# Strong enhancement of *J*<sub>c</sub> in binary and alloyed *in-situ* MgB$_2$ wires by a new approach: Cold high pressure densification

R. Flükiger [o+*], M. S. A. Hossain [+], C. Senatore [o]

[o]Dept. Phys. Mat. Cond. (DPMC), University of Geneva, 1211 Geneva 4, Switzerland

[+]Group of Applied Physics (GAP), University of Geneva, 1211 Geneva 4, Switzerland

[*] Electronic e-mail: Rene.Flukiger@unige.ch

**Abstract**

Cold high pressure densification (CHPD) is presented as a new way to substantially enhance the critical current density of *in situ* MgB$_2$ wires at 4.2 and 20 K at fields between 5 and 14 T. The results on two binary MgB$_2$ wires and an alloyed wire with 10 wt.% B$_4$C are presented The strongest enhancement was measured at 20K, where cold densification at 1.85 GPa on a binary Fe/MgB$_2$ wire raised both $J_c^{//}$ and $J_c^{\perp}$ by more than 300% at 5T, while $B_{irr}$ was enhanced by 0.7 T. At 4.2K, the enhancement of $J_c$ was smaller, but still reached 53% at 10 T. After applying pressures up to 6.5 GPa, the mass density $d_m$ of the unreacted (B+Mg) mixture inside the filaments reached 96% of the theoretical density. After reaction under atmospheric pressure, this corresponds to a highest mass density $d_f$ in the MgB$_2$ filaments of 73%. After reaction, the electrical resistance of wires submitted to cold densification was found to decrease, reflecting an improved connectivity. A quantitative correlation between filament mass density and the physical properties was established.

Monofilamentary rectangular wires with aspect ratios $a/b < 1.25$ based on low energy ball milled powders exhibited very low anisotropy ratios, $\Gamma = J_c^{//}/J_c^{\perp}$ being < 1.4 at 4.2 K and 10T. The present results can be generalized to alloyed $MgB_2$ wires, as demonstrated on a wire with $B_4C$ additives. Based on the present data, it follows that cold densification has the potential of further improving the highest $J_c^{//}$ and $J_c^{\perp}$ values reported so far for *in situ* $MgB_2$ tapes and wires with SiC and C additives. Investigations are under work in our laboratory to determine whether the densification method CHPD can be applied to longer wire or tape lengths.

**Introduction**

The development of $MgB_2$ wires and tapes is in progress and industrial lengths of $MgB_2$ wires and tapes have been produced by both *ex situ* and *in situ* techniques. A recent review article by Collings et al. [1] gives a comprehensive overview of the various attempts which have been undertaken for optimizing $J_c$ in $MgB_2$ wires and tapes. Besides the quality and size of the original powder particles, the strongest raise of $J_c$ has so far been obtained by additives to $MgB_2$. For *in situ* wires Collings et al. [1] have listed not less than 41 additives or additive combinations: it follows that Carbon is always present in the most efficient additives, either as pure C [2,3,4] or in C based compounds, e.g. SiC [5,6], $B_4C$ [7,8], various carbohydrates [9, 10], hydrocarbons [11] and even sugar [12].

Carbon partly substitutes Boron in the $MgB_2$ lattice, thus leading to an enhancement of the electrical resistivity, $\rho_o$, and the irreversible field, $B_{irr}$, and thus $J_c$ at a given field. At the same time $T_c$ decreases gradually, down to < 5K at the solubility limit, x = 0.15 in the formula $Mg(B_{1-x}C_x)_2$ [13]. For round *in situ* wires, the highest $J_c$ value was reported so far by Dou et al. [6] and Susner et al. [14], who found in the superconducting cross section at 4.2 K a value of $J_c = 1 \times 10^4$ A/cm$^2$ at ~ 12.5 T after alloying with 10 wt.% SiC. Recently, the same $J_c$ value was obtained by Hur et al. [15] in the superconducting cross section of a wire produced by an internal Mg diffusion (or infiltration) process.

An enhancement of $J_c$ has been reported by Matsumoto et al. [27] for MgB$_2$ wires with mechanically stronger sheath materials, e.g. stainless steel, which was attributed to an enhanced packing density inside the filaments. The positive effect of stronger sheath materials has also been investigated by Schlachter et al. [28]. In MgB$_2$ tapes, the situation is more complex, due to the anisotropy of the MgB$_2$ phase. The anisotropy ratio $\Gamma = J_c^{//}/J_c^{\perp}$ between the value in parallel magnetic field ($J_c^{//}$) and the one in perpendicular magnetic field ($J_c^{\perp}$) reflects the average *c* axis grain alignment in the filaments (Fujii et al. [16], Kovac et al. [17,18]). After rolling a wire with the initial critical current density $J_c$ to a tape, the latter exhibits the values $J_c^{//}$ and $J_c^{\perp}$, where $J_c^{//} > J_c > J_c^{\perp}$. The difference between $J_c^{//}$ and $J_c^{\perp}$ increases with the degree of texturing, i.e. the higher $J_c^{//}$, the lower $J_c^{\perp}$. The texturing effect is strongest in thin films [19], where the value of $J_c^{//} = 1 \times 10^4$ A/cm$^2$ at 4.2 K is reached at 25 T, the corresponding value of $J_c^{\perp}$ being obtained at a field as low as 10T. An appreciable texturing effect is also observed on *in situ* MgB$_2$ tapes based on mechanically alloyed powders, as shown by Hässler et al. [20]. These authors recently reported a very high value of $J_c^{//} = 1 \times 10^4$ A/cm$^2$ at fields as high as 16.4 T on *in situ* MgB$_2$ tapes with C additives. A rough estimation based on the strong texturing observed in tapes based on mechanically alloyed powders by Kovac et al. [17,18] suggests that the corresponding value of $J_c^{\perp} = 1 \times 10^4$ A/cm$^2$ may be situated below 10 T (see remark [21]).

Another important factor for the critical current density is the connectivity between $MgB_2$ grains, which can be implicitely characterized by the electric resistivity, as stated by Rowell et al. [22]. This property is mainly influenced by the state of interface between neighbor grains but also by the considerable amount of voids in the filaments of *in situ* wires. As mentioned by Collings [1], the mass density inside the filaments of *ex situ* wires is close to 80%, which is considerably higher than for *in situ* wires, where the density does not exceed 50%. The reason for the different densities is linked to the fact that *ex situ* filaments are formed by sintering already formed $MgB_2$ grains, while the *in situ* technique involves a phase reaction process. So far, several attempts have been undertaken in view of the enhancement of the mass density of $MgB_2$ bulk samples and *in situ* filamentary tapes, but all of them are based on a high temperature treatment. Bulk samples were treated at 1400 °C under $\leq 3$ GPa by Lee et al. [13], Karpinski et al. [23], Toulemonde et al. [24]. Prikhna et al. [25] reported at 10K a $J_c$ value of $1 \times 10^4$ A/cm$^2$ at 7.8T for a bulk $MgB_2$ with 10%Ti additive after reaction at 800°C under a pressure of 2 GPa. Reactions of *in situ* wires under HIP conditions (0.2 GPa) were performed by Serquis et al. [26]. Although the resistive measurements reported in Ref. 26 did not cover the whole field range, it can be estimated that at 4.2K, the field at which the value of $J_c = 1 \times 10^4$ A/cm$^2$ was observed was enhanced by approximately 1 T.

Cold high pressure densification is presented here as an alternative way for enhancing the mass density inside $MgB_2$ filaments of wires and tapes, regardless of the sheath material. In the present paper, the efficiency of cold high pressure densification is demonstrated on both binary and $B_4C$ alloyed $MgB_2$ wires. Preliminary results on a series of densified alloyed wires and tapes with malic acid and Carbon additives confirm the present data and will be submitted soon.

**Experimental**

*Sample preparation*

The powders used for the present tapes were Mg from Alfa Aesar (99%, 44 μm) and amorphous Boron (99.9%, average size 1 μm, distribution between 0.05 and 5 μm), while the additive $B_4C$ (size below 50 nm) was supplied by Nanoamor, USA. The powders were mixed by a low energy ball milling process in agate balls and vials, during 1 hour to 4 hours at a rotating speed of 500 rpm in air. No trace of $MgB_2$ was found by X-rays after ball milling. The ball milled powders were filled into pure Fe tubes with an outer diameter of 8 mm and an inner diameter of 5 mm. After packing, the tubes were rotary-swaged to rods of 4.2 mm diameter, then drawn to wires of 1.12 and 0.9 mm diameter as well as of rectangular shapes, after which they were slightly rolled to get a rectangular cross section with low aspect ratio b/a ~ 1.25. The reaction heat treatment was performed under Argon at 1 bar, the treatment for each additive being different, as shown in Table I. The compositions of the powder mixtures in the 3 studied wires was: #1: $MgB_2$, #2: $Mg_{0.9}B_2$ and #3: $MgB_2$ + 10 wt.% $B_4C$.

*Characterization of the wires*

Transport critical current densities $J_c$ were measured as a function of applied magnetic field up to 15 T at $T$ = 4.2 K and 20 K over wire lengths of 45 mm in a He flow cryostat using a four-probe technique, with currents up to 250 A. The temperature was measured on a current lead positioned close to the sample. The voltage taps were 10 mm apart, and the voltage criterion used was 0.1 µV/cm. The irreversibility field $B_{irr}$ was determined using a linear extrapolation of $J_c$ vs. B to 100 A/cm$^2$. The electric resistance was measured in zero field using a dc probe current of 10 mA in the temperature range 10-300 K. The values of $T_c$ were determined from the *R* vs. *T* curves on sheathed filaments. The relative change of the electrical resistance before and after densification has been determined.

The crystal structure analysis was performed by X-ray diffraction, the lattice parameters *a* and *c* being obtained from a Rietveld refinement using the Fullprof software. The microstructure of the filaments was investigated using SEM and optical microscopy.

*Application of pressure*

The pressure on the wires was applied simultaneously from four sides by means of hard metal tools in a prototype cell, resulting in square or rectangular wires. The pressure was applied using a Jossi high precision press with 4 columns on wires having originally either a round or a rectangular cross section, the pressed length being 29 mm. The change of the critical current density was measured on wires pressed up to 4 GPa. The reduction of the mass density of the filaments was determined up to 6.5 GPa at room temperature. The cell design allows a recuperation of the pressed wire without damage. A large number of wires was submitted to a variety of pressures and then characterized. For illustration, the superconducting cross section of some of our wires before and after cold densification is indicated in Table I.

The effect of pressure on the mass density was determined on the basis of volume changes after densification. Wire samples of well defined geometry were measured before and after pressing, the difference in volume being attributed to the (Mg+B) filaments. The change of the filling factor in densified wires was determined by means of SEM and optical microscopy. The uncertainty in the mass density of the (Mg+B) filaments is of the order of 5 %. The inner dimensions of the Fe sheath being essentially unchanged after the reaction heat treatment; this is also the mass density of the reacted $MgB_2$ filaments.

**Results**

*A. Binary $MgB_2$ wires*

*1) Behavior at 4.2K*

The square cross section of wire #1 at the end of the drawing process is shown in Fig. 1a, while Fig. 1b shows the same wire after applying 3 GPa simultaneously on the 4 surfaces, the aspect ratio being now 1: 1.27. The thickness of the reaction zone between $MgB_2$ and the Fe sheath (not shown here) is not affected by the cold densification process.

The application of 1.85 GPa on the binary $MgB_2$ wire #1 caused at 4.2 K an enhancement of $J_c^{//}$ by 52 and 70 % and 8 and 11 T, respectively (Fig. 2). The field at which $J_c$ reaches $1 \times 10^4$ A/cm$^2$ increases from 6.6 to 7.3 T at 4.2 K, i.e. $\Delta B = 0.7$ T. The lattice parameter was not affected by the densification process, the values being $a = 3.0859$ nm, $c = 3.5264$ nm at zero pressure and $a = 3.0851$ nm, $c = 3.5259$ nm after 1.85 GPa.

In Fig. 3, we report the resistivity $\langle \rho \rangle$ for the sample #1 reacted without pressure and after densification at $p = 0.9$, 1.85 and 2.5 GPa. Densification has a negligible effect on the onset $T_c$ and on the transition width, as shown in the inset of Fig. 3. The value of $\langle \rho \rangle$ was determined as $R \cdot S/l$, where $R$ is the measured resistance, $S$ is the cross section of the whole sample and $l$ is the distance between the voltage taps. $\langle\rho\rangle$ thus represents a weighted parallel of the iron and MgB$_2$ resistivity, being

$$\frac{1}{\langle\rho\rangle} = \frac{S_{Fe}}{S}\frac{1}{\rho_{Fe}} + \frac{S_{MgB2}}{S}\frac{1}{\rho_{MgB2}},$$

where $S_{Fe}$ and $S_{MgB2}$ are the cross section areas of iron and MgB$_2$, respectively. The relative value of $\langle\rho\rangle$ was found to decrease with pressure (Fig. 3), indicating an improvement of the grain connectivity.

The results obtained after densification of the stoichiometric binary MgB$_2$ wire #1 were confirmed on a binary wire of the initial composition Mg$_{0.9}$B$_2$ (#2 in Table I). This wire had a square cross section and had an almost isotropic behavior, $J_c^{//}$ and $J_c^{\perp}$ differing by less than 5%. The results at 4.2 K for wire #2 with the initial composition Mg$_{0.9}$B$_2$ are plotted in Fig. 4. After pressing at 3.9 GPa on a square cross section, the critical current density $J_c^{//}$ at 4.2 K and 7 T was found to increase by 65%, while $B_{irr}$ increased from 13.6 to 14.4 T ($\Delta B_{irr}$ = 0.8 T). For comparable pressure values, a similar variation of $J_c^{//}$ and $B_{irr}$ was observed for both wires #1 and #2, thus confirming that the measured effect is due to enhanced connectivity rather than to individual wire parameters. More measurements are needed for a detailed analysis.

The variation of $J_c^{//}$ for the binary MgB$_2$ wire #1 at 7 T after densification is shown in Fig. 5. Only the variation of $J_c^{//}$ will be shown here, the ratio $J_c^{//}/J_c^{\perp}$ being always below 1.3 as a function of applied pressure. Two different behaviors are visible, depending whether the initial shape of the wire before pressing was round or rectangular. For wire

#1 with initially round shape, $J_c^{//}$ increased up to a pressure of 1.8 GPa, where an enhancement by a factor 2 was observed. At pressures > 1.85 GPa an enhancement of $J_c^{//}$ was observed up to 4 GPa for wires with initially rectangular shape, in contrast to round wires, where $J_c^{//}$ vs. $p$ does not show a clear tendency. This different behavior is attributed to the possible formation of cracks in the originally round wires, originating from pressure gradients.

In order to understand the reasons for the enhancement of $J_c^{//}$ and $J_c^{\perp}$ after densification, the change of the overall volume $V = V(Fe) + V(B+Mg)$ of wires with an initial length of 28 mm has been measured. The overall volume reduction $\Delta V$ obtained on the binary MgB$_2$ wire #1 is plotted in Fig. 6 as a function of the applied pressure. All measured samples follow the same tendency, increasing pressures causing a gradual volume reduction. It was found that initially rectangular wires undergo a larger volume reduction than round wires under the effect of the applied pressure. At a pressure of 4 GPa the observed reduction $\Delta V/V$ was 16 and 11 %, respectively.

The question arises about the density change in the unreacted filament under the effect of pressure. From the overall volume reduction $\Delta V$ after densification (Fig. 6), the decrease of the filament volume and thus the enhancement of the (Mg + B) filament mass density $d_m$ in densified wires can be found, the Fe volume being unchanged. As shown in Fig. 7, the variation of mass density $d_m$ of unreacted (Mg+B) filaments reaches 1.98 g/cm$^3$ at 6.5 GPa, thus corresponding to 96% of the theoretical (Mg+B) density.

Since the Fe cross section does not change after the reaction heat treatment, the mass density $d_f$ of the reacted MgB$_2$ filaments in the densified *in situ* wire #1 can now be calculated. The maximum mass density $d_m$ in unreacted (Mg+B) filaments was 1.91

g/cm$^3$ (96% of the theoretical density), which corresponds to a mass density $d_f$ of *in situ* filaments after reaction of 73% of its theoretical density, 2.61 g/cm$^3$. The relative mass density vs. pressure is plotted in Fig. 7 (right scale). It follows that the mass density $d_f$ of MgB$_2$ filaments in the *in situ* wire #1 by cold densification was enhanced from ~42 to ~73% of the theoretical density.

From a comparison between Figs. 3, 5 and 7, it is possible to establish for the first time a quantitative correlation between the observed increase of $J_c$ and of the mass density $d_f$ in the MgB$_2$ filaments as a function of the applied pressure. The observation of a simultaneous decrease of the electrical resistivity confirms that the enhancement of $J_c$ is a direct consequence of the enhanced connectivity in densified wires. A correlated result of cold densification of the present *in situ* wires is the enhancement of the irreversibility field $B_{irr}$, as shown in Table II.

*Behavior at 20K*

The present investigation reveals that the enhancement of $J_c$ of densified wires is considerably higher at 20K than at 4.2 K. After pressing at 1.85 GPa, the value of $J_c^{//}$ at 20K for wire #1 exhibited a strong increase, by 300% in the field range between 4.5 and 5 T, the corresponding enhancement at 4.2K being 53% only (see Fig. 8). At 20K, $B_{irr}^{//}$ in wire #1 after pressing at 1.85 GPa was enhanced by 0.7 T (Table II). This increment $\Delta B_{irr}$ is similar to that at 4.2K, but the relative change $\Delta B_{irr}/B_{irr}$ at 20K is considerably higher than at 4.2K, the corresponding ratios being where 0.12 and 0.05, respectively.

**B. Alloyed MgB$_2$ wires**

In order to get a general answer about the effects of cold densification on *in situ* MgB$_2$ wires, the present investigation was extended to alloyed wires. In the present paper, only the results on a wire containing 10 wt.% B$_4$C (wire #3) are presented. This wire did not exhibit the highest $J_c$ known values [8], which may be due to the quality of the available B$_4$C powder. Nevertheless, the obtained data are representative for illustrating the cold densification effects on alloyed MgB$_2$ wires. MgB$_2$ wires alloyed with other additives, e.g. malic acid and Carbon have also been measured in our laboratory and show even higher enhancements of $J_c$ and $B_{irr}$. The results will be the object of separate publications.

The variation of $J_c$ with applied field for wire #3 at 4.2 K is illustrated in Fig. 9, before and after densification at 1.9 GPa. In a similar way to binary wires, $J_c^{//}$ was found to increase by 40 and 50% at 9 and 11 T, respectively. The irreversibility field $B_{irr}$ was found to increase by 0.7 and 0.8 T, depending whether the initial wire shape before pressing was square or round (Table II). At 4.2K, the alloyed wire #3 after cold densification at 4.2 and at 20 K shows a very similar behavior to that of the binary wires #1 and #2. The corresponding values are listed in Table II.

**Discussion and conclusions**

The critical current density of monofilamentary *in situ* MgB$_2$ wires has been considerably enhanced by using Cold High Pressure Densification. The effects of cold densification have been investigated on two binary wires based on powder mixtures with the nominal compositions MgB$_2$ and Mg$_{0.9}$B$_2$ and on an alloyed wire, the additive being 10 wt.% B$_4$C. The densification procedure was performed at room temperature, just after final wire drawing, at pressures as high as 6.5 GPa.

The present work was performed on *in situ* MgB$_2$ wires based on low energy ball milled precursor powders. The densified wires had rectangular cross sections with very low aspect ratios ($a/b$ = 1.25). The anisotropy ratios $\Gamma = J_c^{//} / J_c^{\perp}$ were well below 1.5 at 4.2K and 10T. The slight change of $\Gamma$ suggests that the degree of texturing was not seriously affected by the densification process.

The main results of the present investigation are the following ones:

*The mass density inside the unreacted (Mg+B) filaments of the binary wire #1 was enhanced up to 96% by the application of pressures up to 6.5 GPa before the final reaction heat treatment. The corresponding mass density enhancement for the alloyed wire was not determined, but the analogy of the behavior of $J_c$ and $B_{irr}$ suggests a similar variation.

*The mass density of the reacted MgB$_2$ filaments was enhanced from 43 up to 73% after cold densification at pressures up to 6.5 GPa.

*The electric resistivity was found to be considerably reduced on densified wires, reflecting an improvement of connectivity between the MgB$_2$ grains.

*A considerable enhancement of the critical current density was observed in all measured densified wires, at 4.2 and 20K in the field range between ~ 5 and 14 T.

  - At 4.2 K, the enhancement of $J_c^{//}$ in the field range between 6 and 12 T was ~ 50%.

  - At 20K, a stronger enhancement was observed, $J_c$ being enhanced by 300% in the field range between 4.5 and 5.5 T after pressing at 1.85 GPa.

*The resistively measured $T_c$ values were not found to change for densified wires. This means that the onset $T_c$ value is unchanged, but does obviously not give any indication about the $T_c$ distribution. Specific heat measurements have been earlier been used for the determination of the $T_c$ distribution in bulk $MgB_2$ samples [29]. Measurements are currently been undertaken for determining the $T_c$ distribution in densified wires.

*After densification at 1.85 GPa an increase of $B_{irr}$ by 0.7 T at 20K was observed for binary $MgB_2$ wires.

In the present paper which was mainly concentrated on binary $MgB_2$ wires, it was found that the improvement at 20K is far more important than at 4.2 K. The results on a wire with $B_4C$ additives are shown for demonstrating that the cold densification effects on $J_c$ are comparable. An even higher enhancement of $J_c$ has been obtained for wires alloyed either with malic acid and Carbon, which were furnished by Matt Rindfleisch (Hyper Tech Research, Inc., Columbus) and by Prof. Y.W. Ma (Institute of Technical Engineering, Beijing), respectively. These results will be published separately. It follows that cold densification has the potential of further improving the highest $J_c^{//}$ and $J_c^{\perp}$ values reported so far for *in situ* $MgB_2$ tapes and wires with SiC and C additives. Investigations are under work in our laboratory to extend the densification method to longer wire or tape lengths.

**Acknowledgments:**

This work was supported by the Swiss National Science Foundation through the National Centre of Competence in Research, Project "Materials with Novel Electronic Properties" (MaNEP/NCCR). The authors would also like to thank F. Buta for useful discussions.

**Table Captions**

**Table I.** Characterization of the monofilamentary rectangular wires studied in the present work. The wire #2 had the nominal composition $Mg_{0.9}B_2$. Reaction conditions: #1 and #2: 1h/650°C, #3: 1h/800°C.

**Table II.** Variation of $B_{irr}$ (defined by $J_c = 10^2$ A/cm$^2$) as a function of applied pressure for the binary Fe/MgB$_2$ wires #1 and #2 and wire #3 with 10 wt.% B$_4$C additive..

**Figure Captions**

**Fig. 1.** Binary Fe/MgB$_2$ wire #1: a) Drawn to a square cross section, b) After cold densification at 3 GPa.

**Fig. 2.** $J_c^{//}$ vs. B at 4.2 K for the Fe/MgB$_2$ wire #1 after densification at 1.5 and 1.85 GPa.

**Fig. 3.** Variation of $\langle\rho\rangle$ for the binary MgB$_2$ wire heat-treated 1h at 650°C at various pressures. The decrease of $\langle\rho\rangle$ with the pressure reflects the improved grain connectivity.

**Fig. 4.** $J_c$ vs. $B$ at 4.2 K for wire #2 (Fe/Mg$_{0.9}$B$_2$) before and after cold densification of initially square wires at 2.35 and 3.90 GPa.

**Fig. 5.** Variation of $J_c^{//}$ of the binary MgB$_2$ wire #1 at 7T and 4.2K as a function of the cold densification pressure. Initially round wires show a maximum of $J_c^{//}$ at 1.8 GPa, while initially rectangular wires still show an enhancement up to 4 GPa. The lines are a guide for the eye.

**Fig. 6.** Overall volume change in the $MgB_2$/Fe wire #1 as a function of applied pressure. The initially rectangular wires showed a higher volume reduction with respect to round wires.

**Fig. 7.** Mass densities of filaments in the $MgB_2$/Fe wire #1. Left scale: relative density of unreacted (Mg+B) mixture; Right scale: relative mass density of reacted $MgB_2$ filaments with respect to the theoretical value.

**Fig. 8.** $J_c^{//}$ and $J_c^{\perp}$ vs. applied field for the binary wire #1 at 4.2 and 20K. Note the considerably higher improvement after pressing at 20 K.

**Fig. 9.** Variation of $J_c^{//}$ vs. B for the *in situ* $MgB_2$ wire alloyed with 10 wt.% $B_4C$ (wire #3) before and after cold densification at 1.25 and 1.9 GPa.

Table. I

| Wire | Additive | Sheath material | $P$ [GPa] | Wire size [mm x mm] | $MgB_2$ area [mm$^2$] | $T_c$ [K] | Fill factor |
|---|---|---|---|---|---|---|---|
| #1 | No additive | Fe | 0 | Ø 1.12 | 0.36 | 37.8 | 0.37 |
| #1 | No additive | Fe | 1.85 | 1.10 X 0.88 | 0.33 | 37.8 | 0.34 |
| #2 | No additive | Fe | 0 | 0.91 X 0.91 | 0.277 | | 0.335 |
| #2 | No additive | Fe | 3.9 | 0.70 X 1.0 | 0.165 | | 0.235 |
| #3 | 10 wt.% B$_4$C | Fe | 0 | Ø 0.90 | 0.216 | 37.1 | 0.34 |
| #3 | 10 wt.% B$_4$C | Fe | 1.9 | 0.99 X 0.62 | 0.196 | | 0.32 |

Table. II

| Wire Sample | $P$ [GPa] | $B_{irr}$ ($T$ =4.2 K) [T] | $B_{irr}$ ($T$ =20 K) [T] |
|---|---|---|---|
| #1   Binary MgB$_2$ (round) | 0 | 12.8 | 6.0 |
|  | 0.9 | 13.3 |  |
|  | 1.85 | 13.3 (//) | 6.7 (//) |
|  | 1.85 | 13.0 ($\perp$) | 6.6 ($\perp$) |
|  | 2.5 | 13.3 |  |
| #2   Binary Mg$_{0.9}$B$_2$ (square) | 0 | 13.6 |  |
|  | 2.35 | 14.4 (//) |  |
|  | 2.35 | 14.2 ($\perp$) |  |
|  | 3.90 | 14.56 |  |
| #3   MgB$_2$+10wt% B$_4$C (round) | 0 | 15.4 |  |
|  | 1.90 | 16.1 |  |
| #3   MgB$_2$+10wt% B$_4$C (square) | 0 | 15.1 |  |
|  | 1.90 | 15.9 |  |

**a)**

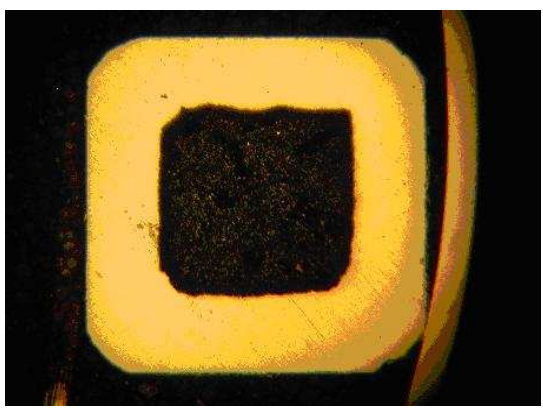

**b)**

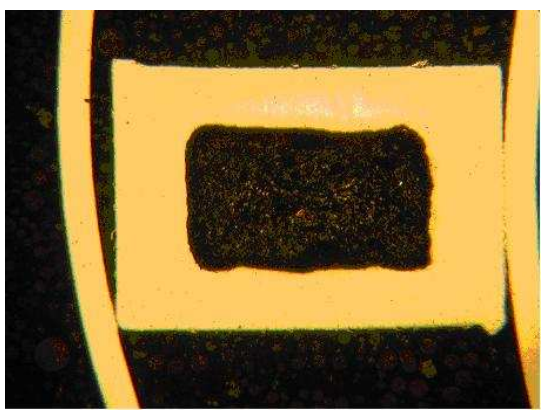

**Fig. 1 Flükiger *et al*.**

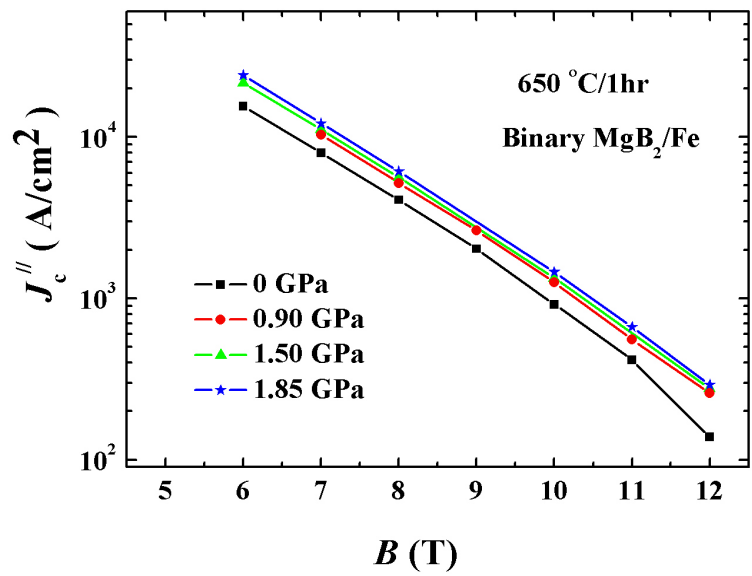

**Fig. 2 Flükiger *et al*.**

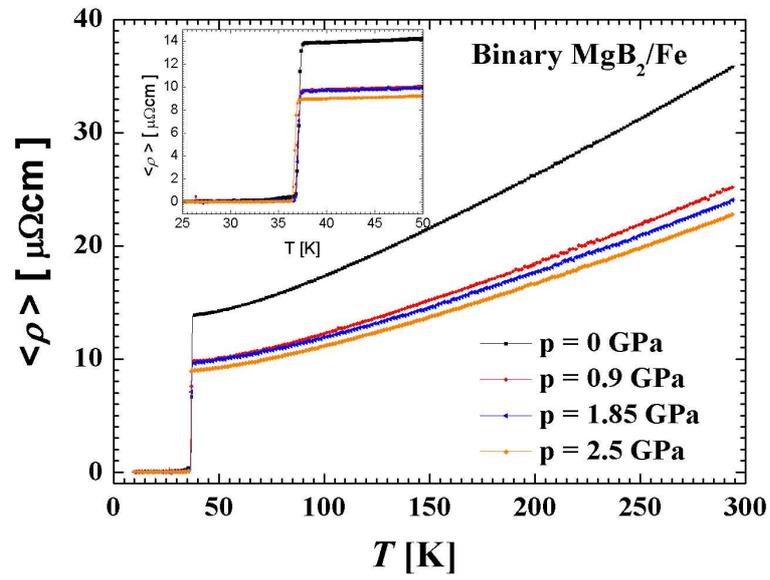

**Fig. 3 Flükiger *et al.*

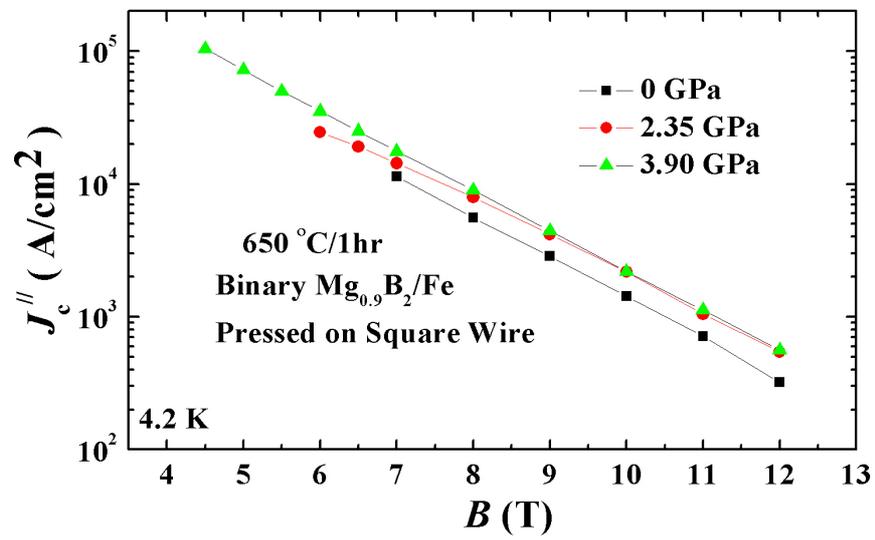

**Fig. 4** Flükiger *et al.*

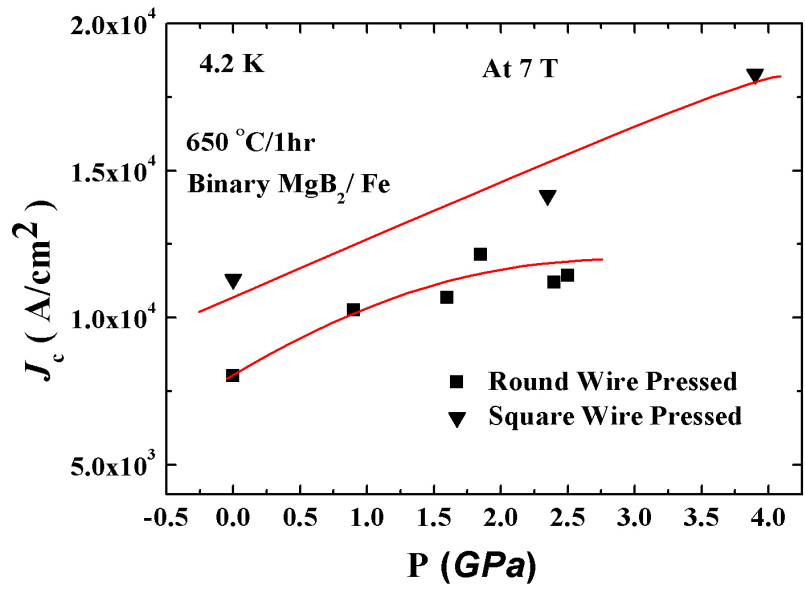

**Fig. 5 Flükiger *et al.***

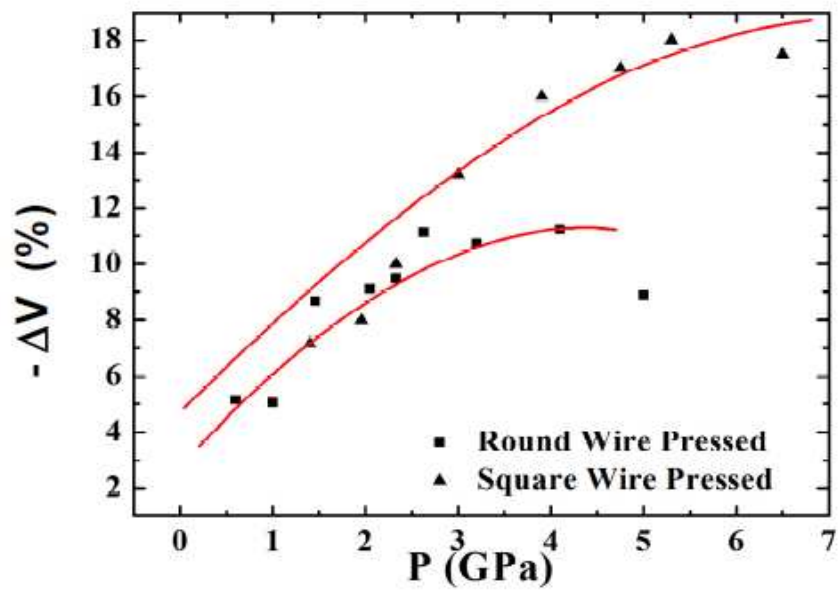

**Fig. 6 Flükiger *et al.***

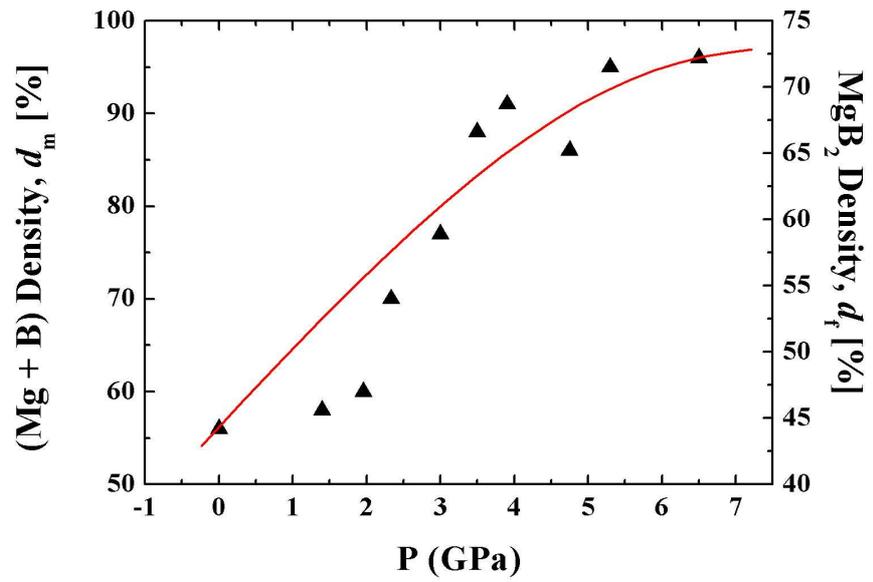

**Fig. 7 Flükiger *et al*.**

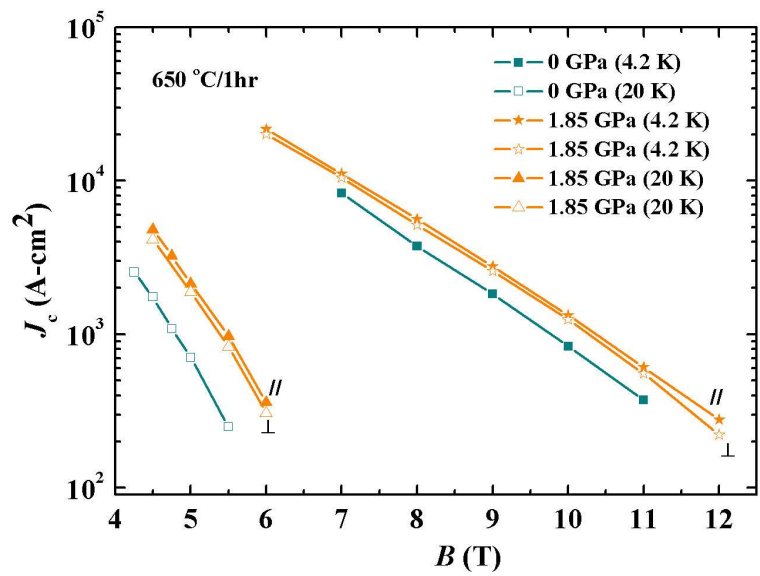

**Fig. 8 Flükiger** *et al.*

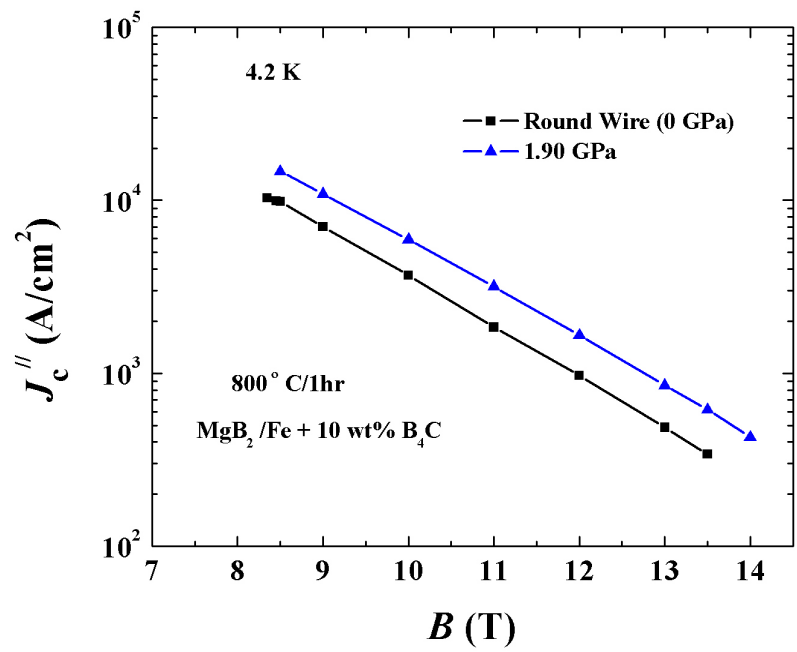

**Fig. 9 Flükiger *et al.***